# Evidence for a 5 MeV Spectral Deviation in the Gösgen Reactor Neutrino Oscillation Experiment

V. Zacek[a]*, G. Zacek[b], P. Vogel[c] and J.-L. Vuilleumier[d]

[a]*Groupe de physique des particules, Université de Montréal, Montréal, H3C 3J7, Canada*
[b]*CresedoTec, Montréal, H2X 2E9, Canada*
[c]*Kellogg Radiation Laboratory, California Institute of Technology, Pasadena, CA 91125, US*
[d]*LHEP, Albert Einstein Center, University of Bern, Switzerland*

Recent reactor neutrino experiments observe an anomalous excess around 5 MeV in the prompt energy spectra of positrons following the inverse beta decay reaction $\bar{\nu}_e + p \rightarrow n + e^+$. In this note we suggest that this 5 MeV anomaly was already present in a much earlier experiment, which was carried out in the 1980's at three distances from the 2.8 GW nuclear powerplant at Gösgen, Switzerland. In particular we demonstrate with a log-likelihood test performed on the Gösgen data that the no – anomaly hypothesis is disfavored at a level of 3.8 sigma.

*Corresponding author: zacekv@lps.umontreal.ca

## 1. Introduction

Four of the recent reactor anti-neutrino experiments, Daya Bay [1], RENO [2], Double Chooz [3] and NEOS [4] observe an unexpected excess of anti-neutrinos with energies between 4.8 and 7.3 MeV. All four experiments are operated in the vicinity of the cores of commercial pressurized water reactors at distances from 24 m up to 1.8 km. The observed excess appears to be independent of the respective distances and has been estimated to be about 1% of all events in both the near and far detectors. The used detection reaction is the inverse beta decay (IBD) $\bar{\nu}_e + p \rightarrow n + e^+$, which has a neutrino energy threshold of $E_{th}$ = 1.8 MeV. All four experiments employ large volumes of Gd - loaded liquid scintillator, where the positron energy is recorded together with the two 511 - keV annihilation gamma rays. Therefore, the visible or prompt energy in the detectors is related to the neutrino energy by $E_{vis} = E_\nu$ - 0.8 MeV and the above-mentioned excess in neutrinos manifests itself as a bump around 5 MeV when plotted as a ratio of data vs. expectation. At present it is not clear what physics gives rise to this so-called "5 MeV bump" [5].

In this note we suggest that the 5 MeV anomaly was already present in a much earlier experiment, which was carried out in the 1980's at three distances from the 2.8 GW nuclear powerplant at Gösgen, Switzerland [6]. However, at that time the slight spectral distortions which showed up at all three distances were not given much consideration, especially since they did not fit any neutrino oscillation hypothesis [7]. In hindsight this is regrettable, because presumably at that time K. Schreckenbach and collaborators [8], who pioneered the measurements of integral beta spectra from fission products at the Institut Laue Langevin (ILL-Grenoble) would have been immediately ready to check for possible origins of the observed anomaly.

## 2. The Gösgen Experiment

In contrast to the recent experiments mentioned above, the Gösgen set up was a segmented detector, approximately one cubic meter in size, consisting of two systems of counters, which recorded, respectively, the positron and the neutron from the IBD reaction. Thirty cells, filled with a total of 377 L of liquid scintillator and arranged in 5 planes served both as a target for incident antineutrinos and as detector for the generated positrons. The reaction neutrons emerging with an energy of several keV were thermalized in the scintillator cell within a few μsec and diffused into one of the adjacent wire chambers filled with $^3$He, where they where detected. To optimize light collection and energy resolution, the scintillator was contained in Lucite cells (6 per plane) with external dimensions of 88 x 20 x 9 cm$^3$ and the individual cells were equipped on both ends with two photomultipliers, coupled together. Pulse shape discrimination was applied to the scintillator signals to distinguish genuine positron events from recoil protons induced by fast background neutrons.

The energy response of the detector to positrons with a given energy and spatial distribution inside a cell was calculated by Monte Carlo simulations and included several effects, such as bremsstrahlung, annihilation at rest and in flight, wall - and escape effects. It is important to notice that, in the given geometry, annihilation at rest deposits little energy in a cell due to the 12 cm long attenuation length of the 511-keV γ rays in the scintillator, leading only to a distortion of the upper flank of the positron peak. As an example, the response function $r(E_{e^+}, E'_{e^+})$ for monoenergetic positrons with $E_{e^+} = 5$ MeV is nearly gaussian, shifted to $E'_{e^+} = 5.16$ MeV with a width of 0.78 MeV FWHM and with a flat tail towards lower energies, essentially due to escape effects (Fig.4 in [6]). This differs from the prompt energy deposition in the recent single volume experiments, which record the entire deposited energy, including that of the annihilation gamma rays.

## 3. The Gösgen Positron Spectra

The energy spectrum of the IBD positrons was measured at three distances (G1, G2, G3) at 37.9, 45.9 and 64.7 m from the reactor core and roughly $10^4$ antineutrinos were registered at each of the three measuring positions. Two analyses were performed in [6], one based on a relative comparison of the measurements at the three distances from the core, while the second analysis compared each spectrum with the positron yield calculated from the expected anti-neutrino spectrum. The neutrino yields for the two dominant isotopes $^{235}$U and $^{239}$Pu were obtained by converting β - spectra which were independently measured at ILL Grenoble with a beta spectrometer using fissioning $^{235}$U and $^{239}$Pu targets [8,9]. The contributions of the less important isotopes $^{238}$U and $^{241}$Pu were obtained from theoretical calculations of the anti-neutrino spectra by Vogel et. al. [10]. No evidence for neutrino oscillations was found at these distances, in agreement with the now well-established oscillation parameters $\Delta m^2_{ee}$ and $\theta_{13}$. Still the data showed a slight upward fluctuation around 4.2 - 4.5 MeV at all three positions with respect to the ILL+Vogel model, corresponding to a range in prompt or visible energy of 5 - 5.3 MeV in the recent experiments.

In order to enhance the statistical significance of an eventual, distance independent spectral deviation from the ILL+Vogel model, the integral G2 and G3 spectra were rescaled by us to the G1 position. The effect of differing burn ups in the three positions was taken into account and contributed a small correction at the level of 0.6% for G2 and 2.4% for G3. The three spectra were then combined by taking the weighted average for each energy bin. The same procedure was applied to the ILL+Vogel predictions for the three positions. The combined spectrum is shown in Fig. 1a and corresponds to an averaged contribution of the fissile isotopes of $^{235}$U (58%), $^{239}$Pu (30%), $^{238}$U (7%) and $^{241}$Pu (5%), respectively. These averaged fission fractions are almost identical to those in the RENO and Daya Bay experiments. The solid curve represents the positron

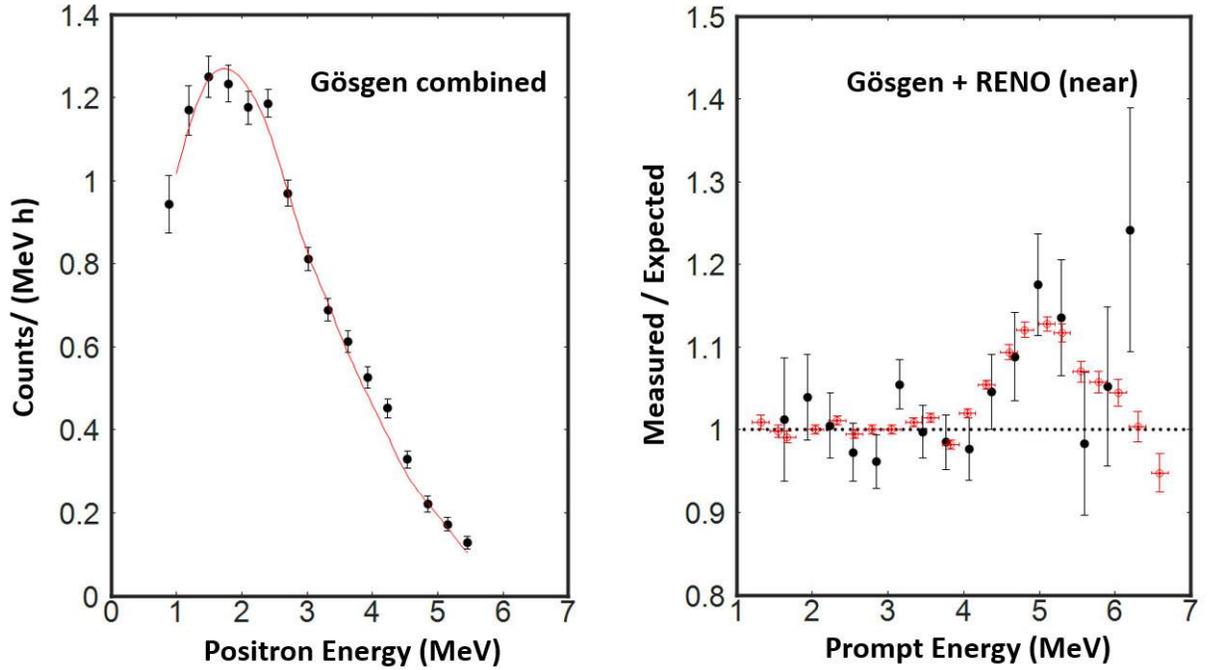

Fig.1 (a) Combined positron spectrum derived from the data taken at 37.9, 45.9 and 64.7 m from the Gösgen PWR core [6]. The G2 and G3 spectra were rescaled to the G1 position. The three spectra were then combined by taking the weighted average of all three energy bins. The bin width is 0.305 MeV and errors shown are statistical. The solid red curve represents the positron yield predicted by the ILL+Vogel model [8-10]. (b) Ratio of the combined measured energy spectrum (black dots) to the predicted spectrum (ILL+Vogel). A shift in the energy scale of 0.84 MeV has been applied to the Gösgen data in order to correct for the incomplete absorption of the two 511-keV annihilation gamma rays. Shown for comparison is the ratio of the RENO near position spectrum [11] to the theoretical prediction by Huber-Mueller [12,13] (red circles).

yield predicted by the ILL+Vogel model. A spectral discrepancy at positron energies around 4.2 MeV is clearly visible.

The ratio of the combined measured energy spectrum to the predicted spectrum (ILL+Vogel) is shown in Fig. 1b. A shift in the energy scale of 0.84 MeV was applied to the Gösgen data in order to match the full energy recorded in recent unsegmented detectors like RENO, Daya Bay, Double Chooz and NEOS [1-4]. Shown for comparison here is the ratio of the RENO near position spectrum [11] to the theoretical prediction of Huber-Mueller [12,13]. The Gösgen data are in good agreement with the respective RENO data. In order to quantify this comparison, the RENO positron spectrum was interpolated to the Gösgen binning, yielding a $\chi^2$ of 12 for 16 degrees of freedom and a corresponding p-value of 0.8. Corrections for the differing energy responses were investigated, but had a negligible effect within the limited statistical accuracy. It is also important to note that the comparison of Daya Bay with Huber Mueller and with the ILL+Vogel model shows the same level of local deviation between 4 and 6 MeV [1].

In order to evaluate the significance of the observed deviation in the Gösgen data we performed a log likelihood ratio test, using the RENO data as a fixed prior to describe the anomaly. The likelihood functions $-lnL_1 = \frac{1}{2}\chi^2(flat)$ and $-lnL_2 = \frac{1}{2}\chi^2(bump)$ compare the ratio data to the null hypothesis and to the interpolated RENO data. With these quantities the log-likelihood ratio $\Delta lnL = \frac{1}{2}[\chi^2(bump) - \chi^2(flat)]$ was calculated. Finally, $10^6$ data sets drawn from the null hypothesis were simulated by Monte Carlo with the combined Gösgen statistics and the resulting log-likelihood ratio distribution was calculated. It turned out that the null - hypothesis is excluded by the data at the level of $1.6 \times 10^{-4}$, that is with a significance of 3.8 σ. It is interesting to note that the reactor spectrum tested at that time by another segmented detector, Bugey - 3 did not observe any spectral distortion when compared to the ILL based prediction within errors of 2.5% in the region of interest [14].

## 4. Conclusions

The energy spectra of positrons from inverse beta decay reactions recorded at three positions of the Gösgen power reactor were statistically combined and compared to predictions (ILL+Vogel). An excess of events was found with a significance of 3.8σ, consistent in energy and amplitude with the flux anomalies recorded by Daya Bay, RENO, Double Chooz and NEOS. In contrast to these recent, large, single volume detectors, Gösgen was a fine grained, segmented experiment, with the prompt signal essentially given by the kinetic energy of the positron only. The presence of the excess in the segmented Gösgen geometry supports its IBD nature and disfavours sterile neutrinos interacting inelastically with $^{13}C$, with subsequent detection of 4.4 MeV de-excitation γ rays, as suggested in [15]. In conclusion, the origin of the so-called "5 MeV bump" remains at the moment unclear and puzzling. Hopefully the situation will be clarified within the next round of very short baseline experiments and at reactors with highly enriched $^{235}U$ cores.

**Acknowledgements**

We would like to thank cordially F. Boehm, who pioneered together with the late R. Mössbauer the Gösgen experiment, for discussions. In addition, we are indebted to all our colleagues and friends of the former Gösgen collaboration.